\documentclass[conference]{IEEEtran}
\usepackage{amsmath,graphicx}
\usepackage{color}
\usepackage{graphicx}
\usepackage{epstopdf}
\usepackage{amsmath}
\usepackage{amssymb}
\usepackage{mathrsfs}
\usepackage{hyperref}
\usepackage{tikz}
\usepackage{bm}
\usepackage[english]{babel}
\usepackage{cite}
\usepackage{rotfloat}
\usepackage{mathtools}
\usepackage[font=normalsize,labelfont=bf]{caption}
\usepackage{amsmath}
\usepackage{makecell}
\usepackage{algorithm,algorithmic}
\usepackage{multirow}
\usepackage{booktabs}
\usepackage{colortbl}
\usepackage{multirow}
\usepackage{hhline}
\usepackage{stfloats}
\usepackage{multicol}
\usepackage{bbm}
\usepackage{cases}
\graphicspath{ {Figures/} }
\newsavebox{\foobox}

\definecolor{kugray5}{RGB}{224,224,224}

\usepackage[normalem]{ulem}
\newcommand\rsout{\bgroup\markoverwith
	{\textcolor{red}{\rule[0.5ex]{2pt}{0.8pt}}}\ULon}



\makeatletter
\newcommand{\ALOOP}[1]{\ALC@it\algorithmicloop\ #1%
	\begin{ALC@loop}}
	\newcommand{\ENDALOOP}{\end{ALC@loop}\ALC@it\algorithmicendloop}

\makeatother

\usepackage{etoolbox}
\let\mybibitem\bibitem
\renewcommand{\bibitem}[1]{%
	\ifstrequal{#1}{nature}
	{\color{blue}\mybibitem{#1}}
	{\color{black}\mybibitem{#1}}%
}

\graphicspath{ {Figures/} }

\DeclareCaptionLabelSeparator{periodspace}{.\quad}

\addto\captionsenglish{}
\allowdisplaybreaks

\usepackage{setspace}

\newcommand{\mH}{{\mathbf{H}}}

\newcommand{\mI}{\textbf{\textbf{I}}}

\newcommand{\mF}{{\mathbf{F}}}

\newcommand{\vs}{{\mathbf{s}}}
\newcommand{\vx}{{\mathbf{x}}}
\newcommand{\vy}{{\mathbf{y}}}

\newcommand{\vn}{{\mathbf{n}}}

\newcommand{\va}{{\mathbf{a}}}

\def\Re{\mathop{\mathtt{Re}}}
\def\Im{\mathop{\mathtt{Im}}}

\def\b0{{\pmb{0}}}

\usepackage{subcaption} 
\newcommand{\mHd}{\mH_{\mathsf{d}}} 
\newcommand{\mHt}{\mH_{\mathsf{t}}} 
\newcommand{\mHr}{\mH_{\mathsf{r}}} 

\begin{document}
%
\title{Deep Reinforcement Learning for Hybrid RIS Assisted MIMO Communications
}
%
%
%


\author{
    \IEEEauthorblockN{Phuong~Nam~Tran\IEEEauthorrefmark{1}, Nhan~Thanh~Nguyen\IEEEauthorrefmark{1}, Markku~Juntti\IEEEauthorrefmark{1},}
    \IEEEauthorblockA{\IEEEauthorrefmark{1}Centre for Wireless Communications, University of Oulu, P.O.Box 4500, FI-90014, Finland
    \\\{phuong.tran, nhan.nguyen, markku.juntti\}@oulu.fi}
}
%
%

\markboth{Journal of \LaTeX\ Class Files,~Vol.~14, No.~8, August~2015}%
{Shell \MakeLowercase{\textit{et al.}}: Bare Demo of IEEEtran.cls for IEEE Journals}
%



\maketitle

\begin{abstract}
    Hybrid reconfigurable intelligent surfaces (HRIS) enhance wireless systems by combining passive reflection with active signal amplification. However, jointly optimizing the transmit beamforming with the HRIS reflection and amplification coefficients to maximize spectral efficiency (SE) is a non-convex problem, and conventional iterative solutions are computationally intensive. To address this, we propose a deep reinforcement learning (DRL) framework that learns a direct mapping from channel state information to the near-optimal transmit beamforming and HRIS configurations. The DRL model is trained offline, after which it can compute the beamforming and HRIS configurations with low complexity and latency.
    Simulation results demonstrate that our DRL-based method achieves 95\% of the SE obtained by the alternating optimization benchmark, while significantly lowering the computational complexity. 
\end{abstract}
\begin{IEEEkeywords}
 hybrid relay-reflecting intelligent surface, MIMO,  proximal policy optimization, spectral efficiency.
\end{IEEEkeywords}

%
\IEEEpeerreviewmaketitle

\section{Introduction and Motivation}
The reconfigurable intelligent surfaces (RISs) have emerged as a key technology for enhancing the coverage and capacity of future wireless networks \cite{liu2021reconfigurable}. The RIS is a planar meta-surface comprising a large number of passive elements, each applying a controllable phase shift to the incident signal to achieve the desired reflection direction \cite{zhou2023survey}. The main advantages of passive RISs are their low cost and high energy efficiency (EE), as they operate without RF chains or active amplification components \cite{zhou2023survey, di2022reconfigurable}. However, their performance is fundamentally limited by the multiplicative fading effect \cite{nguyen2022hybrid, nguyen2022hybridconf}. This results in limited capacity gains, particularly when the direct link is weak or the RIS is far from the transceivers. 

To overcome the limitations of passive RIS, the concept of active RIS was introduced, where each element actively amplifies the incident signals using external power sources \cite{long2021active}. Although active RISs can mitigate severe path loss, it increases hardware complexity and energy consumption. To balance these trade-offs, the hybrid active–passive RIS (HRIS) architecture has been proposed \cite{nguyen2022hybrid, nguyen2022hybridconf}. An HRIS integrates a small number of active elements with many passive ones, leveraging both passive reflection and active amplification capabilities to significantly enhance performance of various wireless systems \cite{nguyen2022hybrid, nguyen2022hybridconf, nguyen2022spectral, nguyen2022hybridUAV, nguyen2022hybridMISO, nguyen2022downlink}.

Maximizing the spectral efficiency (SE) of HRIS-assisted systems requires the joint optimization of the base station (BS) transmit beamforming and the HRIS reflection and amplification coefficients, subject to BS transmit power and HRIS amplification constraints. The unit-modulus constraints on passive elements further introduce non-convexity, rendering the problem high-dimensional and computationally intractable \cite{nguyen2022hybrid, nguyen2022hybridconf, ju2024beamforming}. Existing works rely on iterative methods such as alternating optimization (AO), successive convex approximation (SCA) \cite{nguyen2022hybrid}, block coordinate ascent (BCA) \cite{nguyen2022hybridconf}, majorization–minimization \cite{ju2024beamforming}, and semidefinite programming (SDP) \cite{mu2023efficient}. While these approaches can achieve optimal solutions, their iterative nature results in high complexity and latency, limiting their practicality in dynamic wireless environments.

Recently, deep learning (DL) and deep reinforcement learning (DRL) have become promising tools for handling high-dimensional, non-convex optimization problems in wireless systems. Several works have used supervised learning models, such as convolutional and deep neural networks, to design RIS phase shifts or beamforming for improving data rate and EE \cite{taha2021enabling, alexandropoulos2020phase, ozdogan2020deep}. However, supervised learning relies on large labeled datasets and does not adapt well to changing channel conditions. 
In contrast, DRL learns directly from interactions with the wireless environment, enabling it to adapt to dynamic networks. Prior studies have applied DRL to optimize RIS phase shifts for improving sum rate or data rate \cite{yang2020deep, taha2020deep}, and to jointly optimize BS transmit power and RIS phases for enhancing secrecy rate \cite{yang2020intelligent, guo2021learning}. DRL has also been used to maximize the signal-to-noise ratio (SNR) \cite{feng2020deep}, sum rate \cite{huang2020hybrid}, energy efficiency (EE) \cite{lee2020deep}, and channel capacity \cite{huang2020reconfigurable}. However, these works mainly consider conventional passive RISs in simplified single-user or single-antenna settings and do not address the added complexity of HRIS-assisted multiple-input multiple-output (MIMO) systems, where both reflection and amplification coefficients of the RIS must be jointly optimized.

This paper proposes a novel DRL-based framework that jointly optimizes the BS transmit beamforming and the HRIS reflection and amplification coefficients to maximize the SE of HRIS-assisted MIMO systems. The DRL agent is trained offline to learn an efficient mapping from channel state information (CSI) to near-optimal beamforming vectors and HRIS configurations. This mapping enables fast, low-complexity optimization of BS transmit beamforming and HRIS settings during inference. The framework includes separate training for HRIS setups with fixed active-element positions and for scenarios where the active elements are selected dynamically. Overall, it offers a low-complexity, scalable solution to the high-dimensional and non-convex optimization problem in HRIS-assisted MIMO systems.

\section{System Model and Problem Formulation}
\subsection{System Model}
\begin{figure}[t!]
    \centering          
    \includegraphics[width=0.45\textwidth]{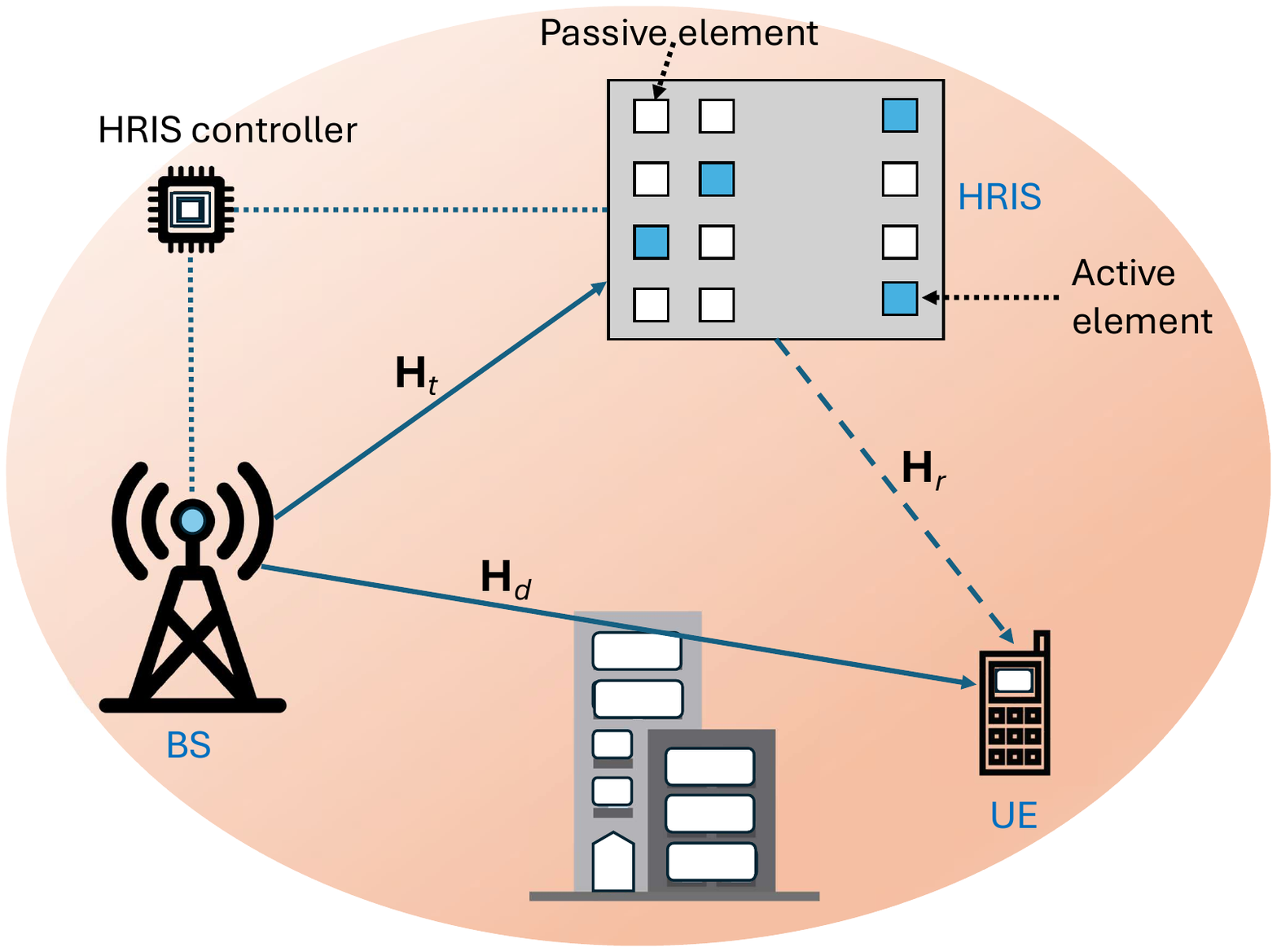}
    \caption{HRIS assisted downlink MIMO system.}
    \label{fig:system}
\end{figure}

We consider a downlink MIMO system where a BS equipped with $N_\mathsf{t}$ antennas serves a user equipped with $N_\mathsf{r}$ antennas, as illustrated in Fig.~\ref{fig:system}. The communication is assisted by an HRIS composed of $N$ elements, among which a subset indexed by $\mathcal{A} \subseteq \{1, \ldots, N\}$ of size $K = |\mathcal{A}|$ is active ($K \ll N$), while the remaining $N - K$ elements operate passively. 
The BS–user, BS–HRIS, and HRIS–user channels are denoted by $\mHd \in \mathbb{C}^{N_\mathsf{r} \times N_\mathsf{t}}$, $\mHt \in \mathbb{C}^{N \times N_\mathsf{t}}$, and $\mHr \in \mathbb{C}^{N_\mathsf{r} \times N}$ , respectively.
The BS transmits a signal vector $\vs = \mF\vx$, where $\vx \in \mathbb{C}^{N_\mathsf{s} \times 1}$ is the vector of $N_\mathsf{s}$ data streams satisfying $\mathbb{E}[\vx\vx^H] = \mI_{N_\mathsf{s}}$, and $\mF \in \mathbb{C}^{N_\mathsf{t} \times N_\mathsf{s}}$ is the transmit precoding matrix. The HRIS modifies the incident signal through a diagonal coefficient matrix $\mathbf{\Theta} = \mathrm{diag}(\alpha_1, \dots, \alpha_N)$, where the reflection coefficient of the $n$-th element is
\begin{equation}
\alpha_n = 
\begin{cases} 
      a_{\text{active}} e^{j\phi_n},  & \text{if } n \in \mathcal{A}, \\
      e^{j\phi_n},      &  \text{otherwise} 
\end{cases}
\end{equation}
where $\phi_n \in [0, 2\pi)$ is the phase shift and $a_{\text{active}} > 1$ is the amplification factor for active elements.
The effective channel combining the direct and HRIS-assisted paths is $\mH_{\mathsf{eff}} = \mHd + \mHr\mathbf{\Theta}\mHt$. The signal received at the user is then given by
\begin{equation}
    \vy = \mH_{\mathsf{eff}}\mF\vx + \vn,
\end{equation}
where the effective noise $\vn$ is the total of three components: the AWGN at the user, $\vn_{u} \sim \mathcal{CN}(\mathbf{0}, \sigma^2\mI_{N_\mathsf{r}})$; the amplified thermal noise from the HRIS active elements, $\vn_{h} \sim \mathcal{CN}(\mathbf{0}, \sigma^2\mI_{N})$; and the amplified residual self-interference (SI) noise, $\vn_{\mathsf{si}} \sim \mathcal{CN}(\mathbf{0}, \eta\sigma^2\mI_{N})$ also at the active elements, where $\eta \geq 0$ denotes the residual SI factor. 
The active coefficient matrix is obtained as $\mathbf{\Theta}_{\mathcal{A}} = \mathbf{J}_{\mathcal{A}}\mathbf{\Theta}$, where $\mathbf{J}_{\mathcal{A}} = \mathrm{diag}(\nu_1, \dots, \nu_N) \in \{0,1\}^{N \times N}$ is a diagonal binary selection matrix whose $n$-th diagonal entry $\nu_n = 1$ if the $n$-th element is active; otherwise, $\nu_n = 0$.
Then, $\vn$ is given as
\begin{equation}
    \vn = \vn_u + \mHr \mathbf{\Theta}_{\mathcal{A}} (\vn_h + \vn_{\mathsf{si}}).
\end{equation}
and its covariance matrix is expressed as \cite{nguyen2022hybrid}
\begin{equation}
    \mathbf{R}_n = \sigma^2 \left( \mI_{N_\mathsf{r}} + (1+\eta) \mHr \mathbf{\Theta}_{\mathcal{A}}\mathbf{\Theta}_{\mathcal{A}}^H \mHr^H \right).
\end{equation}
where the second term captures the amplified noise propagated from the HRIS active elements.

\subsection{Problem Formulation}
Our objective is to maximize the SE of the system by jointly optimizing the BS precoding matrix $\mF$ and the HRIS coefficient matrix $\mathbf{\Theta}$. The SE (in bps/Hz) is expressed as
\begin{equation}
    R(\mF, \mathbf{\Theta}) = \log_2 \left| \left(\mI_{N_\mathsf{r}} + \frac{1}{\sigma^2} \mH_{\mathsf{eff}}\mF\mF^H\mH_{\mathsf{eff}}^H \mathbf{R}_n^{-1} \right) \right|.
\end{equation}
The optimization problem is formulated as
\begin{subequations}
\begin{align}
\underset{\mF, \mathbf{\Theta}}{\text{maximize}} \quad & R(\mF, \mathbf{\Theta}) \label{eq:obj} \\
\text{subject to} \quad & ||\mF||_F^2 \le P_{\text{max}}^{\text{BS}}, \label{eq:bs_power} \\
& |\alpha_n| = 1, \quad \forall n \notin \mathcal{A}, \label{eq:passive_const} \\
& |\alpha_n| = a_{\text{active}}, \quad \forall n \in \mathcal{A}, \label{eq:active_amp}\\
& 0 \le \phi_n < 2\pi, \quad \forall n = 1, \dots, N. \label{eq:phase_const}
\end{align}
\end{subequations}
where $P_{\text{max}}^{\text{BS}}$ denotes  the maximum BS transmit power.
Constraint \eqref{eq:bs_power} limits the BS transmit power, while \eqref{eq:passive_const} and \eqref{eq:active_amp} enforce the unit-modulus condition on passive elements and control the amplification of active ones, respectively. Constraint \eqref{eq:phase_const} specifies the range of phase shifters.
The objective function $R(\mF, \mathbf{\Theta})$ is non-convex with respect to $\mF$ and $\mathbf{\Theta}$. Consequently, the problem \eqref{eq:obj} is non-convex and high-dimensional due to the large number of RIS elements, making global optimization computationally intractable. This motivates the development of a low-complexity solution based on DRL, as elaborated next.

\section{Proposed DRL Approach}
\subsection{DRL Framework}
In this section, we propose a DRL framework that learns to jointly optimize the transmit beamforming and HRIS configurations, with low computational complexity. The SE maximization problem is modeled as a reinforcement learning (RL) task \cite{sutton1998reinforcement}, in which a DRL agent deployed at the BS learns an efficient mapping from instantaneous CSI to the corresponding beamforming and HRIS coefficients. The key components of the proposed RL framework include state, action, and reward, which are defined as follows.

\textbf{1) State:} 
At each time step $t$, the agent observes a state $\vs_t \in \mathcal{S}$, where $\mathcal{S}$ represents the state space derived from the instantaneous channel information. To fully characterize the wireless environment, the state $\vs_t$ is defined based on the complete CSI as
\begin{equation}
\begin{aligned}
    \vs_t &= [\text{vec}(\Re(\mHd^{(t)})); \text{vec}(\Im(\mHd^{(t)})); \\
        &\quad \text{vec}(\Re(\mHt^{(t)})); \text{vec}(\Im(\mHt^{(t)})); \\
        &\quad \text{vec}(\Re(\mHr^{(t)})); \text{vec}(\Im(\mHr^{(t)}))] \in \mathbb{R}^{D_s},
\end{aligned}
\end{equation}
where $\mHd^{(t)}$, $\mHt^{(t)}$, and $\mHr^{(t)}$ represent the BS–user, BS–HRIS, and HRIS–user channel matrices at time $t$, respectively. The operators $\Re(\cdot)$ and $\Im(\cdot)$ extract the real and imaginary components, and $\text{vec}(\cdot)$ stacks a matrix into a vector. Consequently, the state dimension is $D_s = 2(N_\mathsf{r} N_\mathsf{t} + N N_\mathsf{t} + N_\mathsf{r} N)$.

\textbf{2) Action:} 
The action $\va_t \in \mathcal{A}$ corresponds to the optimization variables in \eqref{eq:obj}, including the BS precoding matrix $\mF^{(t)}$ and the HRIS amplification and phase-shift coefficients $\{a_n^{(t)}\}$ and $\{\phi_n^{(t)}\}$, where $\mathcal{A}$ denotes the action space. It is defined as
\begin{equation}
\begin{aligned}
    \va_t &= [\text{vec}(\Re(\mF^{(t)})); \text{vec}(\Im(\mF^{(t)})); \\
          &\quad a_1^{(t)}, \dots, a_N^{(t)}; \phi_1^{(t)}, \dots, \phi_N^{(t)}] \in \mathbb{R}^{2N_\mathsf{t} N_\mathsf{s} + 2N},
\end{aligned}
\end{equation}

\textbf{3) Reward:} 
To match the objective in \eqref{eq:obj}, the reward is defined as the SE achieved by executing action $\va_t$  in state $\vs_t$,
\begin{equation}
    r_t = R(\mF^{(t)}, \mathbf{\Theta}^{(t)}),
\end{equation}
where $\mF^{(t)}$ and $\mathbf{\Theta}^{(t)}$ are derived from $\va_t$. 
Maximizing $r_t$ encourages the agent to select configurations that improve SE.

The objective of the DRL agent is to learn an optimal deterministic policy $\pi^*$ that maps each state $\vs_t$ to an action $\va_t = \pi(\vs_t)$. The policy $\pi$ is implemented as a deep neural network with parameters $\Psi_{\pi}$. The optimal parameters $\Psi_{\pi}^*$ are obtained by maximizing the expected reward over the channel state distribution $\mathcal{D}$, formulated as
\begin{equation}
    \Psi_{\pi}^* = \arg \max_{\Psi_{\pi}} \mathbb{E}_{\vs_t \sim \mathcal{D}} \big[ r(\vs_t, \pi_{\Psi_{\pi}}(\vs_t)) \big].
    \label{eq:RL_problem}
\end{equation}
Once trained, the policy network performs inference with low computational complexity, offering an efficient solution for joint transmit beamforming and HRIS optimization.

\subsection{Proximal Policy Optimization}
To solve the formulated RL problem, we employ the proximal policy optimization (PPO) algorithm \cite{schulman2017proximal}, an on-policy actor-critic method known for stable training and strong performance in high-dimensional continuous control. PPO mitigates unstable policy updates using a clipped surrogate objective, making it suitable for our high-dimensional continuous action space. The agent consists of a policy network (actor) $\pi_{\Psi}$ and a value network (critic) $V_{\Phi}$, which are trained using data collected from interactions with the wireless environment. 
At each training step $t$, the actor takes the state $\vs_t$ as input and outputs a probability distribution over the action space, from which an action $\va_t$ is sampled. The agent then executes $\va_t$ and receives a reward $r_t$, while the critic estimates the value of $\vs_t$. 
Parameters $\Psi$ and $\Phi$ are updated based on batches of collected experiences to improve both the policy and value estimation.

The parameter update begins with computing the temporal-difference (TD) error
\begin{equation}
    \delta_t = r_t + \gamma V_{\Phi}(\vs_{t+1}) - V_{\Phi}(\vs_t),
\end{equation}
where $\gamma \in [0,1]$ is the discount factor. The advantage is then estimated using generalized advantage estimation (GAE) \cite{schulman2015high}, given by
\begin{equation}
    \hat{A}_t = \sum_{k=0}^{T-1} (\gamma\lambda)^k \delta_{t+k},
\end{equation}
where $\lambda \in [0,1]$ is the GAE trade-off parameter, and $T$ is the batch length.
The critic aims to improve the state-value estimation using the target value, defined as
\begin{equation}
    V_t^{\text{targ}} = \hat{A}_t + V_{\Psi_{\text{old}}}(\vs_t).
\end{equation}
where $V_{\Psi_{\text{old}}}$ is the critic before the update. 
The critic's parameters are then updated by minimizing the mean-squared error loss $L^{V}(\phi) = \mathbb{E}_{t} [ (V_{\Phi}(\vs_t) - V_t^{\text{targ}})^2 ]$ via gradient descent
\begin{equation}
    \Phi \leftarrow \Phi - \alpha_V \nabla_{\Phi} L^{V}(\Phi),
\end{equation}
where $\alpha_V$ is the critic learning rate.

Simultaneously, the actor is updated by maximizing the PPO clipped objective, defined as
\begin{equation}
    L^{\text{CLIP}}(\Psi) = \mathbb{E}_{t} \left[ \min\left( p_t(\Psi) \hat{A}_t, \text{clip}(p_t(\Psi), 1-\epsilon, 1+\epsilon) \hat{A}_t \right) \right],
\end{equation}
where $p_t(\Psi) = \frac{\pi_{\Psi}(\va_t|\vs_t)}{\pi_{\Psi{\text{old}}}(\va_t|\vs_t)}$, $\Psi{\text{old}}$ are the previous policy parameters, and $\epsilon$ is the clipping threshold. The actor parameters are updated via gradient ascent, given by
\begin{equation}
    \Psi \leftarrow \Psi + \alpha_{\pi} \nabla_{\Psi} L^{\text{CLIP}}(\Psi).
\end{equation}
where $\alpha_{\pi}$ is the actor learning rate.
This training process iteratively collects trajectories, computes advantages, and updates both networks over multiple epochs until convergence.

\subsection{PPO-Based Algorithm}
\begin{figure}
  \centering          
  \includegraphics[width=0.45\textwidth]{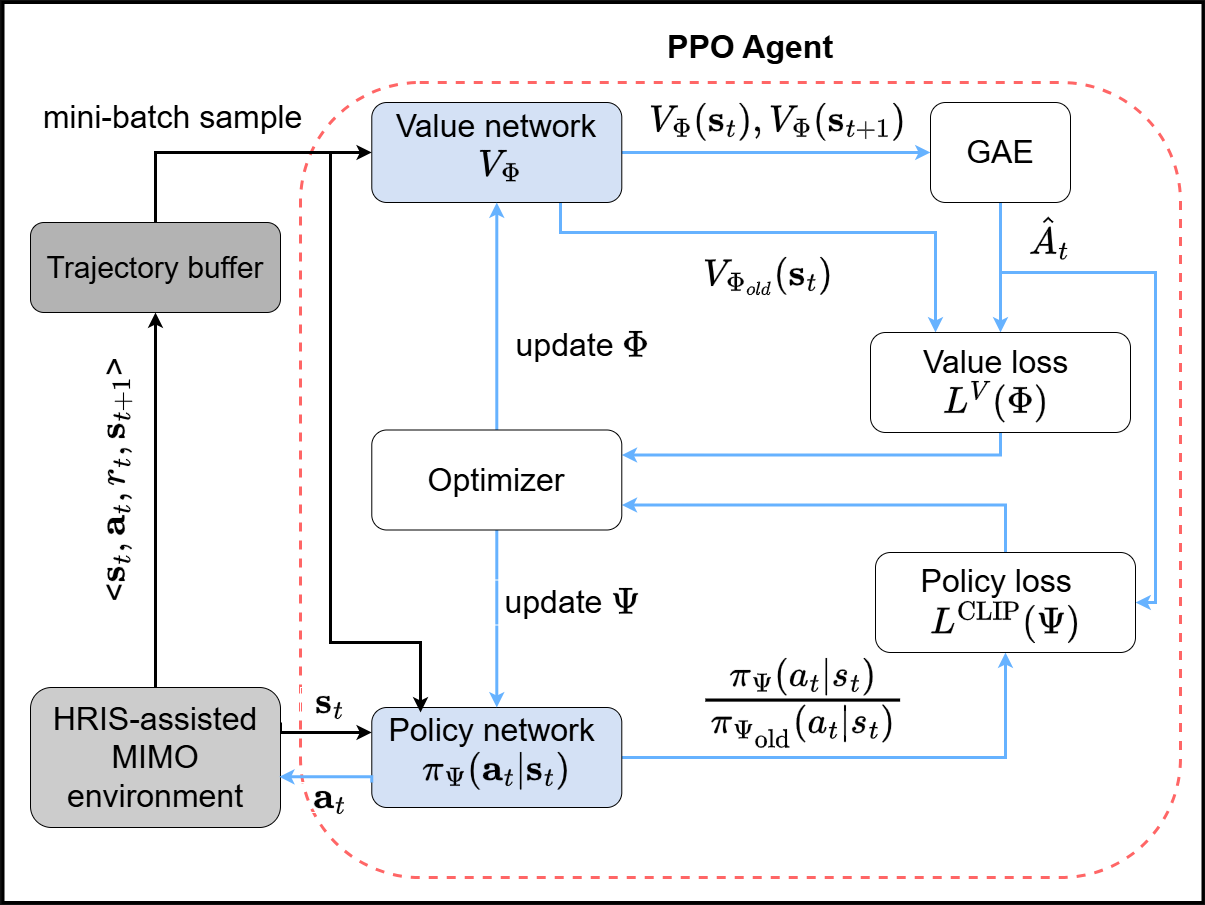}
  \caption{The proposed DRL framework.}
  \label{fig:ppo_workflow}
\end{figure}
Figure~\ref{fig:ppo_workflow} illustrates the training process of the proposed PPO-based algorithm for jointly optimizing BS transmit beamforming and HRIS reflection and amplification coefficients.
At each interaction step, the policy network $\pi_{\Psi}$ observes the current state $\vs_t$ from the HRIS-assisted MIMO environment and generates an action $\va_t$, which specifies the BS beamforming and HRIS configuration. The environment then returns the next state $\vs_{t+1}$ and the immediate reward $r_t$, and the transition tuple $\langle \vs_t, \va_t, r_t, \vs_{t+1} \rangle$ is stored in a trajectory buffer. 
After collecting a batch of trajectories, mini-batches are sampled to compute the critic’s value estimates $V_{\Phi}(\vs_t)$ and $V_{\Phi}(\vs_{t+1})$, and the advantage estimates $\hat{A}_t$ using GAE. 
These are then used to update the value network by minimizing the value loss $L^{V}(\Phi)$ and to update the policy network by maximizing the PPO clipped surrogate objective $L^{\text{CLIP}}(\Psi)$. 
This iterative process of trajectory collection, advantage computation, and policy/value updates continues until convergence, enabling the agent to learn a stable and efficient policy for joint beamforming and HRIS optimization.

\section{Simulation Results}
\begin{figure*}[t!]
    \centering
    \begin{subfigure}[b]{0.32\textwidth}
        \centering
        \includegraphics[width=\textwidth]{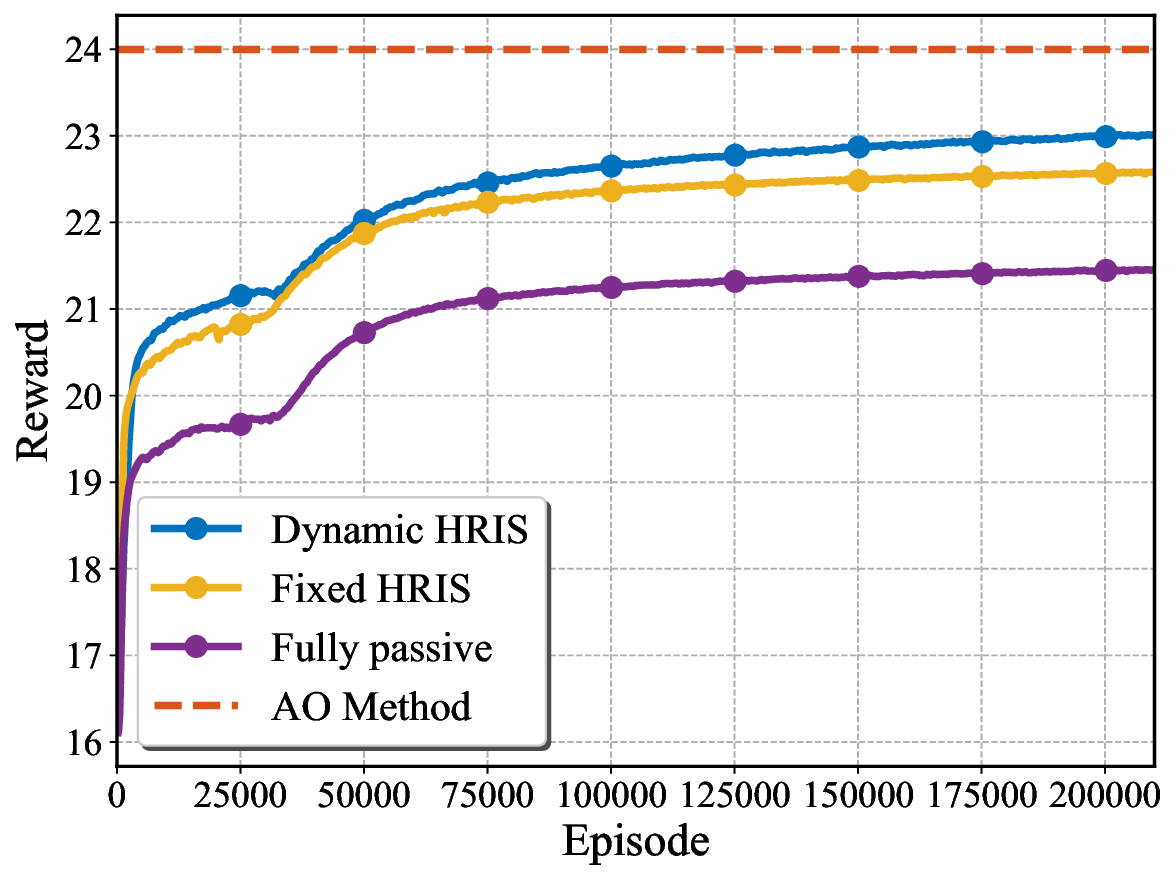}
        \caption{Convergence of the training reward.}
        \label{fig:convergence}
    \end{subfigure}
    \hfill
    \begin{subfigure}[b]{0.32\textwidth}
        \centering
        \includegraphics[width=\textwidth]{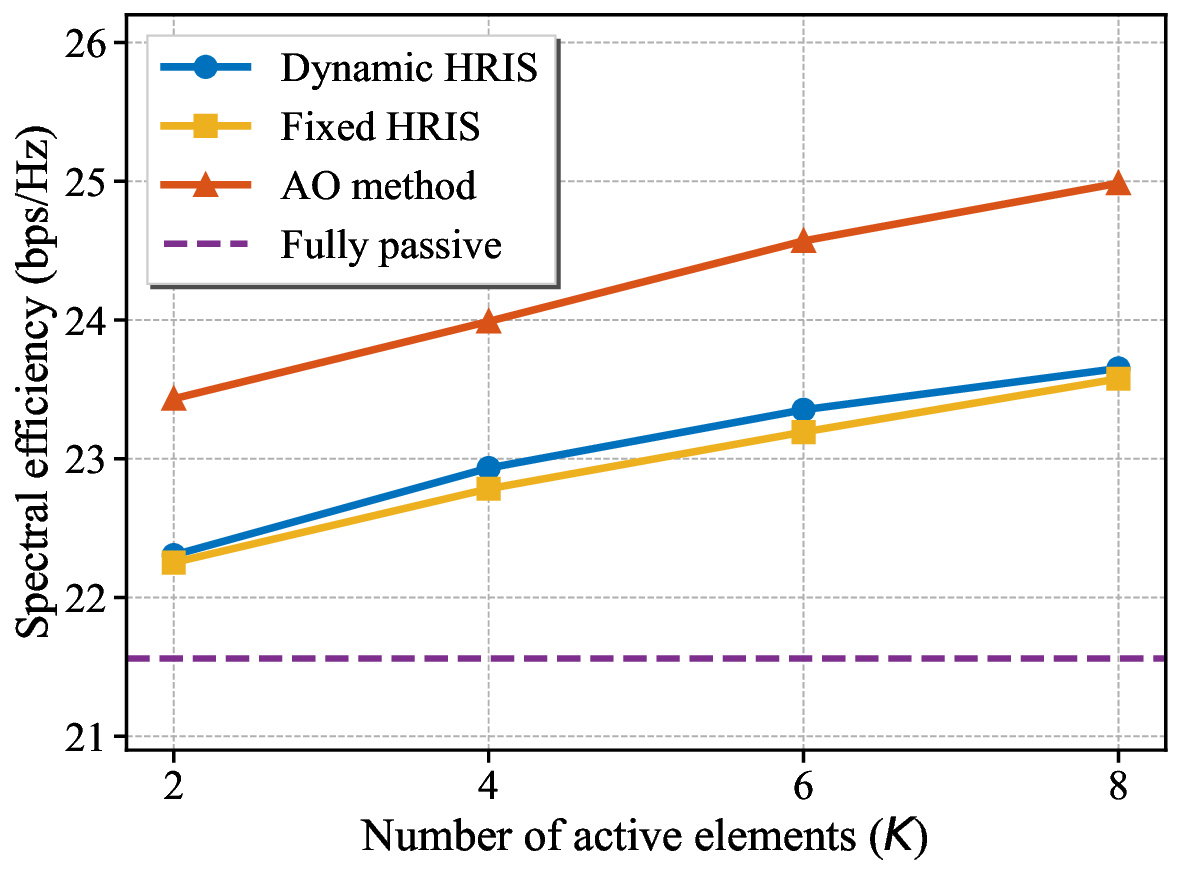}
        \caption{Spectral efficiency versus number of active elements.}
        \label{fig:SE_vs_active_elm}
    \end{subfigure}
    \hfill
    \begin{subfigure}[b]{0.32\textwidth}
        \centering
        \includegraphics[width=\textwidth]{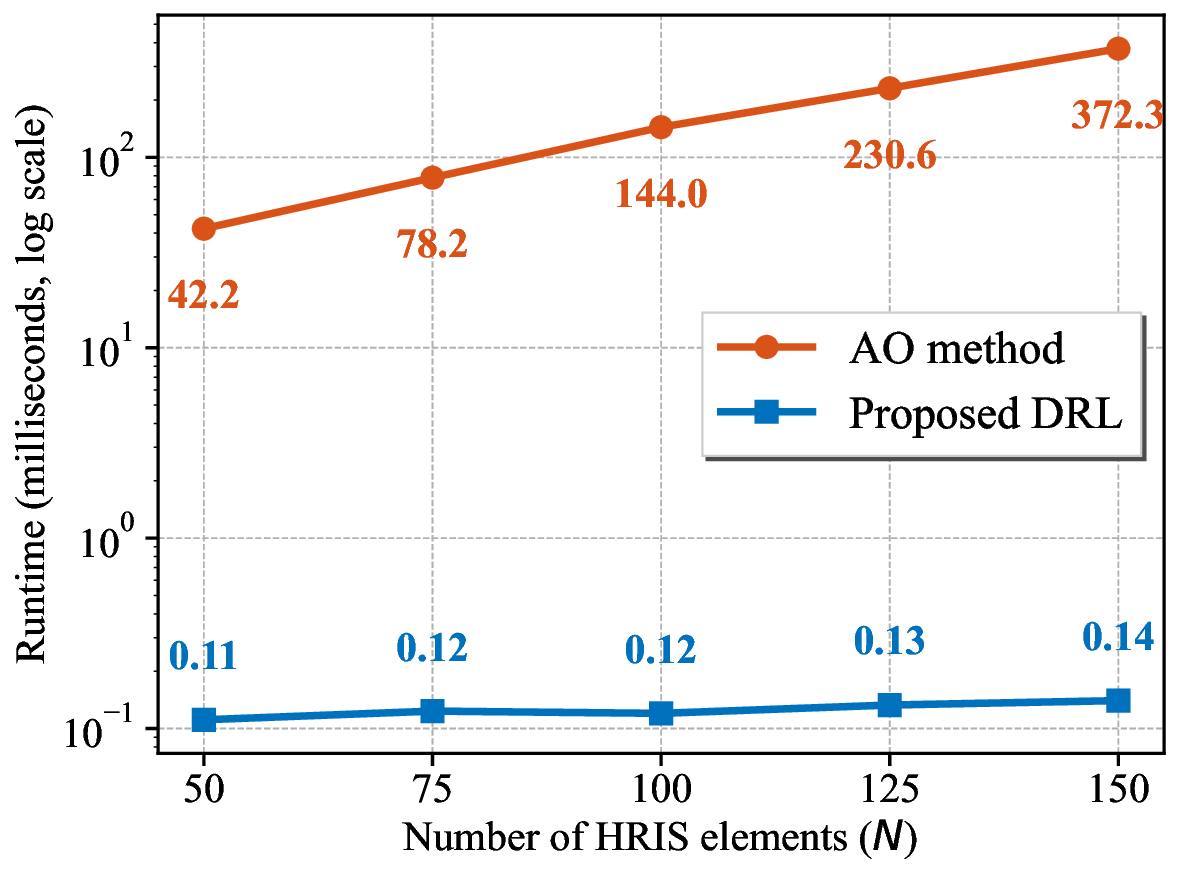}
        \caption{Average runtime versus the number of HRIS elements.}
        \label{fig:runtime}
    \end{subfigure}
    \caption{Convergence, spectral efficiency, and runtime performance of the proposed DRL scheme compared with benchmarks.}
    \label{fig:all_results}
\end{figure*}

We consider a downlink MIMO scenario where the BS, HRIS, and user are located at $(0,0)$~m, $(50,0)$~m, and $(45,2)$~m, respectively.
The large-scale fading follows the path loss model $\beta(d) = \beta_0 (d/d_0)^{-\epsilon}$, 
where $\beta_0 = -30$~dB at $d_0=1$~m \cite{zhang2020capacity}. The path loss exponents for the BS–user, BS–HRIS, and HRIS–user links are set to $\{3.5,\,2.2,\,2.0\}$, respectively.
All wireless channels follow a Rician fading model, where the Rician factors $\kappa_d$, $\kappa_t$, and $\kappa_r$ correspond to the BS–user, BS–HRIS, and HRIS–user links, following the model in \cite{nguyen2022hybrid}. We set $\{\kappa_\mathsf{d}, \kappa_\mathsf{t}, \kappa_\mathsf{r}\} = \{0, 1, \infty\}$, indicating Rayleigh fading on the BS–user link, equal LoS and NLoS power on the BS–HRIS link, and a dominant LoS component on the HRIS–user link.
The system bandwidth is $20$~MHz, and the noise power is $\sigma^2 = -169~\text{dBm/Hz} + 10\log_{10}(20 \times 10^6) + 10~\text{dB}$, corresponding to a noise figure of $10$~dB. The normalized residual self-interference is set to $\eta = 1$~dB. The BS and user are equipped with $N_\mathsf{t} = 4$ and $N_\mathsf{r} = 2$ antennas, respectively. The HRIS comprises $N = 50$ elements, with $K$ active elements varied to to evaluate system performance. The maximum amplification factor is $a_{\text{active}} = 10$ \cite{saikia2025hybrid}. The BS transmit power is $P_{\text{BS}} = 40$~dBm.

The DRL agent is trained using the PPO algorithm. The actor and critic networks each consist of two hidden layers with 256 neurons and employ rectified linear unit (ReLU) activation functions. Training is performed using the Adam optimizer with a learning rate of $\alpha_{\pi} = \alpha_V = 0.001$. The PPO clipping parameter is set to $\epsilon = 0.2$, while the discount factor and GAE parameter are chosen as $\gamma = 0.99$ and $\lambda = 0.95$, respectively. The agent is trained for 200{,}000 episodes, each consisting of 1000 time steps, and the network parameters are updated using batches of 2048 collected samples.

To comprehensively evaluate the proposed DRL-based approach, we consider two HRIS configurations: the \textbf{dynamic HRIS} and the \textbf{fixed HRIS}. In the dynamic configuration, the DRL agent jointly optimizes the BS precoder, the HRIS coefficients, and the selection indices of $K$ active elements in each channel realization. In the fixed configuration, the set of $K$ active elements is predetermined, allowing us to isolate the performance gain introduced by dynamic element selection.
For benchmarking, we include two baselines: (i) a \textbf{fully passive RIS}, where all $N$ elements are passive, and both the BS beamforming and RIS phase shifts are optimized by the PPO agent; and (ii) the \textbf{AO-based optimization} from \cite{nguyen2022hybrid}, which serves as an upper-performance bound.

Fig.~\ref{fig:convergence} shows the training reward curves for the DRL-based schemes under different RIS architectures, considering an HRIS configuration with $K=4$ active elements. The fully passive RIS converges the fastest because its action space is the smallest, involving only passive phase shifts and BS beamforming. 
The dynamic HRIS has the slowest convergence since it must jointly optimize the BS precoder, HRIS coefficients, and the active-element selection. Nevertheless, it achieves the highest final reward among the DRL-based methods, demonstrating the benefit of adaptive element selection. While the AO method still yields the highest performance, the gap between AO and the DRL-based schemes is small, indicating that the proposed approach attains near-optimal performance.

Fig.~\ref{fig:SE_vs_active_elm} shows the SE of the proposed DRL-based schemes as the number of active HRIS elements $K$ increases. All HRIS-assisted configurations improve with larger $K$ because additional active elements enhance signal amplification. The dynamic HRIS consistently outperforms the fixed HRIS, though the gap is small. For example, at $K=6$, the dynamic HRIS reaches 23.36~bps/Hz, slightly higher than the 23.19~bps/Hz achieved by the fixed HRIS. In comparison, the AO method for dynamic HRIS achieves 24.57~bps/Hz, indicating that the DRL-based dynamic and fixed HRIS achieve about $95\%$ and $94\%$ of this performance, respectively. Both DRL-based schemes also outperform the fully passive RIS, demonstrating the advantage of integrating even a few active elements. Overall, although AO yields the highest SE, the DRL-based methods deliver performance close to this upper bound.

Fig.~\ref{fig:runtime} compares the runtime of the proposed DRL framework with that of the iterative AO method, evaluated with $K=4$ active elements. The runtime measures the average time needed to compute the beamforming and HRIS configurations for a channel realization as the number of HRIS elements $N$ varies. The AO method shows a rapid increase in runtime, from 42.2 ms at $N=50$ to 372.3 ms at $N=150$, indicating poor scalability with respect to $N$. In contrast, the inference time of the proposed DRL framework remains nearly constant at 0.11--0.14 ms across all values of $N$. At $N=150$, this yields a speedup exceeding 2500$\times$ over AO. These results confirm that the DRL-based approach achieves near-AO performance while providing significantly lower complexity and enhanced scalability, making it well-suited for real-time dynamic wireless environments.

\section{Conclusion}
In this paper, we proposed a novel DRL-based framework for jointly optimizing the BS beamforming and the HRIS reflection and amplification coefficients in HRIS-aided MIMO systems. To address the inherent non-convexity and high computational complexity of the joint optimization problem, we developed a DRL framework based on the PPO algorithm that learns a direct mapping from the CSI to near-optimal BS and HRIS configurations. Simulation results showed that the proposed framework achieves over 95\% of the SE obtained by the iterative AO baseline while providing a substantial reduction in computational complexity.

\section*{Acknowledgment}
This work was supported in part by European Union through MSCA Doctoral Network EXACT-6G (GA 101120297).

\ifCLASSOPTIONcaptionsoff
  \newpage
\fi



%
\bibliographystyle{IEEEtran}
\bibliography{bibtex/bib/IEEEabrv, bibtex/bib/references}

@STRING{IEEE_J_VT         = "{IEEE} Trans. Veh. Technol."}

@STRING{IEEE_J_COML       = "{IEEE} Commun. Lett."}

@STRING{IEEE_J_JSAC       = "{IEEE} J. Sel. Areas Commun."}

@STRING{IEEE_J_COM        = "{IEEE} Trans. Commun."}

@STRING{IEEE_J_WCOM       = "{IEEE} Trans. Wireless Commun."}

@STRING{IEEE_J_WCOML      = "{IEEE} Wireless Commun. Lett."}

@STRING{IEEE_J_PROC       = "Proc. {IEEE}"}

@STRING{IEEE_O_ACC        = "{IEEE} Access"}

@STRING{IEEE_O_CSTO       = "{IEEE} Commun. Surveys Tuts."}

@string{ globecom = {Proc. IEEE Global Commun. Conf.}}

@string{ icc = {Proc. IEEE Int. Conf. Commun.}}

@string{ spawc = {Proc. IEEE Works. on Sign. Proc. Adv. in Wirel. Comms.}}

@inproceedings{nguyen2022hybridUAV,
  title={Hybrid active-passive reconfigurable intelligent surface-assisted {UAV} communications},
  author={Nguyen, Nhan T and Nguyen, V-Dinh and Wu, Qingqing and T{\"o}lli, Antti and Chatzinotas, Symeon and Juntti, Markku},
  booktitle=globecom,
  pages={3126--3131},
  year={2022}
}

@inproceedings{nguyen2022hybridMISO,
  title={Hybrid active-passive reconfigurable intelligent surface-assisted multi-user {MISO} systems},
  author={Nguyen, Nhan T and Nguyen, V-Dinh and Wu, Qingqing and T{\"o}lli, Antti and Chatzinotas, Symeon and Juntti, Markku},
  booktitle=spawc,
  year={2022}
}

@inproceedings{nguyen2022downlink,
  title={Downlink throughput of cell-free massive {MIMO} systems assisted by hybrid relay-reflecting intelligent surfaces},
  author={Nguyen, Nhan T and Nguyen, V and Nguyen, Hieu V and Ngo, Hien Q and Chatzinotas, Symeon and Juntti, Markku and others},
  booktitle=icc,
  year={2022},
}

@article{liu2021reconfigurable,
  title={Reconfigurable intelligent surfaces: Principles and opportunities},
  author={Liu, Yuanwei and Liu, Xiao and Mu, Xidong and Hou, Tianwei and Xu, Jiaqi and Di Renzo, Marco and Al-Dhahir, Naofal},
  journal=IEEE_O_CSTO,
  volume={23},
  number={3},
  pages={1546--1577},
  year={2021},
  publisher={IEEE}
}

@article{zhou2023survey,
  title={A survey on model-based, heuristic, and machine learning optimization approaches in RIS-aided wireless networks},
  author={Zhou, Hao and Erol-Kantarci, Melike and Liu, Yuanwei and Poor, H Vincent},
  journal=IEEE_O_CSTO,
  volume={26},
  number={2},
  pages={781--823},
  year={2023},
  publisher={IEEE}
}

@article{di2022reconfigurable,
  title={Reconfigurable intelligent surfaces [scanning the issue]},
  author={Di Renzo, Marco and Tretyakov, Sergei},
  journal=IEEE_J_PROC,
  volume={110},
  number={9},
  pages={1159--1163},
  year={2022},
  publisher={IEEE}
}

@article{ju2024beamforming,
  title={Beamforming optimization for hybrid active-passive RIS assisted wireless communications: A rate-maximization perspective},
  author={Ju, Yue and Gong, Shiqi and Liu, Heng and Xing, Chengwen and An, Jianping and Li, Yonghui},
  journal=IEEE_J_COM,
  year={2024},
  publisher={IEEE}
}

@article{nguyen2022spectral,
  title={Spectral efficiency analysis of hybrid relay-reflecting intelligent surface-assisted cell-free massive MIMO systems},
  author={Nguyen, Nhan Thanh and Nguyen, Van-Dinh and Van Nguyen, Hieu and Ngo, Hien Quoc and Chatzinotas, Symeon and Juntti, Markku},
  journal={IEEE Transactions on Wireless Communications},
  volume={22},
  number={5},
  pages={3397--3416},
  year={2022},
  publisher={IEEE}
}

@article{nguyen2022hybrid,
  title={Hybrid relay-reflecting intelligent surface-assisted wireless communications},
  author={Nguyen, Nhan Thanh and Vu, Quang-Doanh and Lee, Kyungchun and Juntti, Markku},
  journal=IEEE_J_VT,
  volume={71},
  number={6},
  pages={6228--6244},
  year={2022},
  publisher={IEEE}
}

@inproceedings{nguyen2022hybridconf,
  title={Hybrid active-passive reconfigurable intelligent surface-assisted multi-user MISO systems},
  author={Nguyen, Nhan T and Nguyen, V-Dinh and Wu, Qingqing and T{\"o}lli, Antti and Chatzinotas, Symeon and Juntti, Markku},
  booktitle=spawc,
  pages={1--5},
  year={2022},
  organization={IEEE}
}

@article{mu2023efficient,
  title={Efficient active elements selection algorithm for hybrid RIS-assisted D2D communication system},
  author={Mu, Gaoze and Zhang, Peichang and Hou, Yanzhao and Zhong, Shida and Huang, Lei and Yuan, Tao},
  journal=IEEE_J_COML,
  volume={28},
  number={2},
  pages={377--381},
  year={2023},
  publisher={IEEE}
}

@article{taha2021enabling,
  title={Enabling large intelligent surfaces with compressive sensing and deep learning},
  author={Taha, Abdelrahman and Alrabeiah, Muhammad and Alkhateeb, Ahmed},
  journal=IEEE_O_ACC,
  volume={9},
  pages={44304--44321},
  year={2021},
  publisher={IEEE}
}

@inproceedings{alexandropoulos2020phase,
  title={Phase configuration learning in wireless networks with multiple reconfigurable intelligent surfaces},
  author={Alexandropoulos, George C and Samarakoon, Sumudu and Bennis, Mehdi and Debbah, M{\'e}rouane},
  booktitle=globecom,
  pages={1--6},
  year={2020},
  organization={IEEE}
}

@article{ozdogan2020deep,
  title={Deep learning-based phase reconfiguration for intelligent reflecting surfaces},
  author={{\"O}zdogan, {\"O}zgecan and Bj{\"o}rnson, Emil},
  journal={arXiv preprint arXiv:2009.13988},
  year={2020}
}

@inproceedings{taha2020deep,
  title={Deep reinforcement learning for intelligent reflecting surfaces: Towards standalone operation},
  author={Taha, Abdelrahman and Zhang, Yu and Mismar, Faris B and Alkhateeb, Ahmed},
  booktitle=spawc,
  pages={1--5},
  year={2020},
  organization={IEEE}
}

@article{yang2020intelligent,
  title={Intelligent reflecting surface assisted anti-jamming communications: A fast reinforcement learning approach},
  author={Yang, Helin and Xiong, Zehui and Zhao, Jun and Niyato, Dusit and Wu, Qingqing and Poor, H Vincent and Tornatore, Massimo},
  journal=IEEE_J_WCOM,
  volume={20},
  number={3},
  pages={1963--1974},
  year={2020},
  publisher={IEEE}
}

@article{guo2021learning,
  title={Learning-based robust and secure transmission for reconfigurable intelligent surface aided millimeter wave UAV communications},
  author={Guo, Xufeng and Chen, Yuanbin and Wang, Ying},
  journal=IEEE_J_WCOML,
  volume={10},
  number={8},
  pages={1795--1799},
  year={2021},
  publisher={IEEE}
}

@article{yang2020deep,
  title={Deep reinforcement learning-based intelligent reflecting surface for secure wireless communications},
  author={Yang, Helin and Xiong, Zehui and Zhao, Jun and Niyato, Dusit and Xiao, Liang and Wu, Qingqing},
  journal=IEEE_J_WCOM,
  volume={20},
  number={1},
  pages={375--388},
  year={2020},
  publisher={IEEE}
}

@article{feng2020deep,
  title={Deep reinforcement learning based intelligent reflecting surface optimization for MISO communication systems},
  author={Feng, Keming and Wang, Qisheng and Li, Xiao and Wen, Chao-Kai},
  journal=IEEE_J_WCOML,
  volume={9},
  number={5},
  pages={745--749},
  year={2020},
  publisher={IEEE}
}

@inproceedings{huang2020hybrid,
  title={Hybrid beamforming for RIS-empowered multi-hop terahertz communications: A DRL-based method},
  author={Huang, Chongwen and Yang, Zhaohui and Alexandropoulos, George C and Xiong, Kai and Wei, Li and Yuen, Chau and Zhang, Zhaoyang},
  booktitle=globecom,
  pages={1--6},
  year={2020},
  organization={IEEE}
}

@inproceedings{lee2020deep,
  title={Deep reinforcement learning for energy-efficient networking with reconfigurable intelligent surfaces},
  author={Lee, Gilsoo and Jung, Minchae and Kasgari, Ali Taleb Zadeh and Saad, Walid and Bennis, Mehdi},
  booktitle=icc}

@article{huang2020reconfigurable,
  title={Reconfigurable intelligent surface assisted multiuser MISO systems exploiting deep reinforcement learning},
  author={Huang, Chongwen and Mo, Ronghong and Yuen, Chau},
  journal=IEEE_J_JSAC,
  volume={38},
  number={8},
  pages={1839--1850},
  year={2020},
  publisher={IEEE}
}

@book{sutton1998reinforcement,
  title={Reinforcement learning: An introduction},
  author={Sutton, Richard S and Barto, Andrew G and others},
  volume={1},
  number={1},
  year={1998},
  publisher={MIT press Cambridge}
}

@article{schulman2017proximal,
  title={Proximal policy optimization algorithms},
  author={Schulman, John and Wolski, Filip and Dhariwal, Prafulla and Radford, Alec and Klimov, Oleg},
  journal={arXiv preprint arXiv:1707.06347},
  year={2017}
}

@article{schulman2015high,
  title={High-dimensional continuous control using generalized advantage estimation},
  author={Schulman, John and Moritz, Philipp and Levine, Sergey and Jordan, Michael and Abbeel, Pieter},
  journal={arXiv preprint arXiv:1506.02438},
  year={2015}
}

@article{zhang2020capacity,
  title={Capacity characterization for intelligent reflecting surface aided MIMO communication},
  author={Zhang, Shuowen and Zhang, Rui},
  journal=IEEE_J_JSAC,
  volume={38},
  number={8},
  pages={1823--1838},
  year={2020},
  publisher={IEEE}
}

@article{saikia2025hybrid,
  title={Hybrid-RIS Empowered UAV-Assisted ISAC Systems: Transfer Learning-based DRL},
  author={Saikia, Prajwalita and Jee, Anand and Singh, Keshav and Huang, Wan-Jen and Boulogeorgos, Alexandros-Apostolos A and Tsiftsis, Theodoros A},
  journal=IEEE_J_COM,
  year={2025},
  publisher={IEEE}
}

@article{long2021active,
  title={Active reconfigurable intelligent surface-aided wireless communications},
  author={Long, Ruizhe and Liang, Ying-Chang and Pei, Yiyang and Larsson, Erik G},
  journal=IEEE_J_WCOM,
  volume={20},
  number={8},
  pages={4962--4975},
  year={2021},
  publisher={IEEE}
}
%








\end{document}